\documentstyle[12pt,psfig]{article}
\begin{document}
\begin{titlepage}
\begin{center}

{\Large\bf{The AGL Equation from the  Dipole Picture}}
\\[5.0ex]
{\Large\it{ M. B. Gay  Ducati $^{*}$\footnotetext{$^{*}$E-mail:gay@if.ufrgs.br}}}\\
 {\it and}\\
{ \Large \it{ V. P.  Gon\c{c}alves $^{**}$\footnotetext{$^{**}$E-mail:barros@if.ufrgs.br} 
}} \\[1.5ex]
{\it Instituto de F\'{\i}sica, Univ. Federal do Rio Grande do Sul}\\
{\it Caixa Postal 15051, 91501-970 Porto Alegre, RS, BRAZIL}\\[5.0ex]
\end{center}

{\large \bf Abstract:}
The AGL equation includes all multiple pomeron exchanges in the double 
logarithmic approximation (DLA) limit, leading to an  unitarized gluon 
distribution in the small $x$ regime. This equation was originally obtained 
using the Glauber-Mueller approach. We demonstrate in this paper that the 
AGL equation and, consequently,  the GLR equation, can also  be obtained  from the   dipole picture in the double logarithmic limit, using an  evolution equation, recently proposed, which includes all multiple pomeron exchanges in 
the leading logarithmic approximation.    Our conclusion is that the AGL equation is 
a good candidate for an unitarized evolution equation at small $x$ in the DLA 
limit.

\vspace{1.5cm}



\end{titlepage}

\section{Introduction}
\label{int}

The physics of high-density QCD has  become an increasingly active subject 
of research, both from  experimental  and  theoretical points of view. 
Presently, and in the near future, the collider facilities such as the DESY 
$ep$ collider HERA ($ep$, $eA$), Fermilab Tevatron ($p\overline{p}$, $pA$), 
BNL Relativistic Heavy Ion Collider (RHIC), and CERN Large Hadron Collider 
(LHC) ($p\overline{p}$, $AA$) will be able to probe new regimes of dense 
quark matter at very small Bjorken $x$ or/and at large $A$, with rather 
different dynamical properties. The description of these processes  is 
directly associated with a correct description of the dynamics in this 
kinematical region.

The behavior of the cross sections in the high energy limit ($s \rightarrow \infty$) and fixed momentum transfer is expected to be described by the BFKL equation \cite{bfkl}. The simplest process where  this equation applies  is the high energy scattering between two heavy quark-antiquark states, {\it i.e.} the onium-onium  scattering. For a sufficiently heavy onium state, high energy scattering is  a perturbative process since the onium radius gives the essential scale at which                                                the running coupling $\alpha_s$ is evaluated. This process was studied in the dipole picture \cite{mueller, mueller1, 
mueller2},  where the 
heavy quark-antiquark pair and the 
soft gluons in the limit of large number of colors  $N_c$  are viewed as a collection of color 
dipoles. In this case, the cross section can be understood as a product of 
the number of dipoles in one onium state times the number of dipoles in the 
other onium state times the  basic cross section for dipole-dipole 
scattering due to two-gluon exchange. The cross section  grows rapidly with the energy ($\sigma \propto \alpha_s^2  \, e^{(\alpha_P - 1)Y}$, where $(\alpha_P - 1) = \frac{4\alpha_s\,N_c}{\pi}\,ln2$ and $Y = ln\,s/Q^2$) because the number of dipoles in the light cone wavefunction grows rapidly with the energy. In \cite{mueller} Mueller demonstrated that the 
QCD dipole picture reproduces the  BFKL physics.

One of the main characteristics of the BFKL equation is that it predicts 
very high density of partons  in the small $x$ region.
 Therefore a new dynamical effect  associated with the unitarity corrections is expected to stop further growth of the parton densities. The understanding of the unitarity corrections has been a challenge of perturbative QCD (PQCD). 

About seventeen years ago, 
Gribov, Levin, and Ryskin (GLR) \cite{100} performed a detailed study of this problem in the double logarithmic approximation (DLA). 
They  argued that  the physical processes of interaction and recombination of 
partons become important in the parton cascade at a large value of the parton 
density, and that these shadowing corrections could be expressed in a new evolution equation - the GLR equation. This equation considers the leading non-ladder contributions:
the multi-ladder diagrams, denoted as fan diagrams. The main characteristics of this 
equation are that  it predicts a saturation of the gluon distribution at very small 
$x$,  it predicts a critical line, separating the perturbative regime from the 
saturation regime, and  it is only valid in the border of this critical line. Therefore,
the GLR equation predicts the limit of its validity.
In the last decade,  the solution \cite{collins,bartels1,bartels2} and 
possible generalizations  \cite{bartels,laenen} of the GLR equation have 
been studied in great detail. Recently, an eikonal approach to  the  
shadowing corrections was proposed in the literature \cite{ayala1,ayala2,ayala3}. 
 The starting point of these papers is the proof of the Glauber formula in QCD \cite{muegla}, which considers only the interaction  of the fastest partons  with the target.
In \cite{ayala2}, a generalized equation which takes into account the interaction 
of all partons in a parton cascade with the target in the DLA limit  was proposed by Ayala, Gay Ducati, and Levin (AGL). 
The main properties of the  generalized equation  are that (i) the iterations of this 
equation coincide with the iteration of the Glauber-Mueller formula; 
(ii) its solution matches  the solution of the DGLAP evolution equation 
in the DLA limit of PQCD; (iii) it 
has the GLR equation as a  limit, and (iv)  contains the 
Glauber-Mueller formula. Therefore, the AGL equation is valid in a 
large kinematic region.

  The AGL equation resums all multiple pomeron exchanges in the DLA limit. Its  assintotic solution is given by $xG \propto Q^2\,R^2\,ln\,(1/x)$, where $R$ is the size of the target,  {\it i.e.} differently from the GLR equation, it does not  predict saturation of  the gluon distribution  in the very small $x$ limit. 

The unitarity corrections in the leading logarithmic limit can be studied using the dipole picture.
In this picture the unitarity corrections (multiple interactions between the onia) become important at high energies \cite{mueuni,salam}. These corrections in general are neglected as being of higher order in $1/N_c$. However, when the energy is high, there is a large number of dipoles in each onium, and the total number of possible interactions is of order $e^{(\alpha_P - 1)(y  + (Y-y))}$ (the product of the number of dipoles in each onium) \cite{salam}. So when  $\alpha_s^2 \, e^{(\alpha_P - 1)Y} \approx \, 1$, one should take into account the multiple scattering corrections - multiple pomeron exchange - despite the fact that they are suppressed by $1/N_c^2$. Recently,  an equation which includes all pomeron exchanges in the leading logarithmic approximation using the dipole picture was proposed (Eq. (15) in Ref. \cite{kov}). In this paper we will denote this equation as K equation. As the  K equation and the AGL equation resums the same group of diagrams, both should be identical in a common limit. Our goal in this paper  is to demonstrate that the K equation reproduces    the AGL equation and, consequently,  the GLR equation in the DLA limit.

This paper is organized as follows. In Section 2 the AGL approach for the unitarity corrections is briefly reviewed. We clarify some steps and obtain the GLR equation as a limit case. In  Section 
3 we obtain  the nuclear structure function $F_2^A$ in the Glauber-Mueller approach and  compare it with the expression proposed in \cite{kov}. It allows to obtain precisely the relation between the  propagator of the $q\overline{q}$ through the  nucleus and the gluon distribution. In Section 4, we obtain the AGL equation from the dipole picture and, as a limit, the GLR equation. Finally, in Section 5,  we present 
our  conclusions.

\section{The AGL Equation}
\label{agls}

 In  the nucleus rest frame we can consider the interaction between a  virtual colorless hard probe and the nucleus via a gluon pair ($gg$) component of the virtual probe. In the region where 
$x << 1/2mR$ ($R$ is the size of the target), the $gg$ pair 
crosses the target with fixed
transverse distance $r_t$ between the gluons. 
Moreover, at high energies the lifetime of the $gg$ pair  may substantially exceed the nuclear radius. The cross section for this process is written as \cite{ayala1}
\begin{eqnarray}
\sigma^{G^*A}=  \int_0^1 dz \int \frac{d^2r_t}{\pi} 
 |\Psi_t^{G^*}(Q^2,r_t,x,z)|^2 \sigma^{gg+A}(z,r_t^2)\,\,,
\label{sig1}
\end{eqnarray}
where $G^*$ is the  virtual colorless hard probe with virtuality  $Q^2$, $z$ is the fraction of energy carried by the gluon and $\Psi_t^{G^*}$ is the wave function of the transverse polarized gluon in the virtual probe. Furthermore, $\sigma^{gg+A}(z,r_t^2)$ is the cross section of the interaction of the $gg$ pair with the  nucleus.

 As our goal in this paper is to obtain the AGL from the dipole picture, 
we should make a transformation of the above  equation (\ref{sig1}) for 
$q\overline{q}$ dipoles, since in this picture  they are the 
basic configuration. Considering that $\sigma^{gg + A} = (C_A/C_F) 
\sigma^{q \overline{q}}$ we have that the cross section for the interaction 
between the virtual probe $G^*$ and the nucleus via the $q \overline{q}$ component
of the virtual probe is given by
\begin{eqnarray}
\sigma^{G^*A}= \frac{C_A}{C_F} \int_0^1 dz \int \frac{d^2r_t}{\pi} 
 |\Psi_t^{G^*}(Q^2,r_t,x,z)|^2 \sigma^{q\overline{q}+A}(z,r_t^2)\,\,.
\label{sig2}
\end{eqnarray}

To estimate the unitarity corrections we have to take into account 
the rescatterings of the quark-antiquark  pair inside the nucleus. 
The contributions of the  rescatterings can be estimated  using the 
Glauber-Mueller approach proposed  in  ref. \cite{ayala1}. 
Following the same steps used in \cite{ayala1} , {\it i. e.} considering 
the $s$-channel unitarity and the eikonal model, one 
 obtains that the $\sigma^{G^*A}$ cross section is written as
\begin{eqnarray}
\sigma^{G^*A}= \frac{C_A}{C_F}  \int_0^1 dz \int \frac{d^2r_t}{\pi} 
\int \frac{d^2b_t}{\pi} |\Psi_t^{G^*}(Q^2,r_t,z)|^2 \,2\,
[1 - e^{-\frac{1}{2}\sigma_N^{q\overline{q}}(x^{\prime}
,\frac{4}{r_t^2})S(b_t)}]\,\,, 
\label{sig3}
\end{eqnarray}
where $x^{\prime} = x/(z\,r_t^2\,Q^2)$ ($x$ is the Bjorken variable), $ b_t$ is the impact parameter,  
$S(b_t) = (A/ \pi R^2) e^{-\frac{b_t^2}{R^2}}$ is the gaussian profile 
function  and $\sigma_N^{q\overline{q}}$ is the cross section of the 
interaction of the $q\overline{q}$ pair with the  nucleons inside  the 
nucleus. In \cite{plb} the authors have shown  that $ \sigma_N^{q\overline{q}} 
= \frac{C_F}{C_A} (3 \alpha_s(\frac{4}{r_t^2})/4)\,\pi^2\,r_t^2\,
 xG_N(x,\frac{4}{r_t^2})$, where  $xG_N(x,\frac{4}{r_t^2})$ is the nucleon gluon 
 distribution.

The wavefunction $\Psi^{G^*}$ was calculated in \cite{muegla,ayala1} using the 
technique of Ref. \cite{brodsky}. Here we only explicitate the result for the 
squared wavefunction, which is given by
\begin{eqnarray}
|\Psi_t^{G^*}(Q^2,r_t,z)|^2 = \frac{1}{z(1-z)}\left[ \left(\epsilon^2 K_2(\epsilon r_t) 
- \frac{\epsilon K_1(\epsilon r_t)}{r_t}\right)^2 + \frac{1}{r_t^2} (\epsilon K_1(\epsilon r_t))^2 
\right] \,\,,
\label{wf}
\end{eqnarray}
where $\epsilon^2 = Q^2 z(1-z)$ and the $K_i$ are the modified Bessel functions.
The main contribution in expression (\ref{sig3}) comes from the region of small $z$, 
where $\epsilon r_t \ll 1$. Using the expansion of the modified Bessel functions 
for small values of the argument, results that the squared wavefunction simplifies to
\begin{eqnarray}
|\Psi_t^{G^*}(Q^2,r_t,x,z)|^2 = \frac{2}{z r_t^4}\,\,.
\label{wf2}
\end{eqnarray}
The condition  $\epsilon r_t \ll 1$ implies that 
\begin{eqnarray}
z(1-z) < \frac{1}{Q^2 r_t^2}  < \frac{1}{4}   \,\,,
\label{ine}
\end{eqnarray}
where the last inequality comes from symmetric pairs ($z = 1/2$). From the  
expression (\ref{ine}) results that   $x^{\prime} \ge x$.

The  relation $\sigma^{G^*A}(x,Q^2) = (4\pi^2 \alpha_s/Q^2)xG_A(x,Q^2)$ is valid for a virtual probe $G^*$ with virtuality $Q^2$.  
Consequently, using  the expression (\ref{wf2}) of the squared  wavefunction 
and making the   
change of variables $z \rightarrow x^{\prime}$, we obtain that   
the Glauber-Mueller formula for the interaction of the $q\overline{q}$ pair with the 
nucleus is written as
\begin{eqnarray}
xG_A(x,Q^2) = \frac{4}{\pi^2}  \frac{C_A}{C_F} \int_x^1 
\frac{dx^{\prime}}{x^{\prime}} \int_{\frac{4}{Q^2}}^{\infty} \frac{d^2r_t}{\pi r_t^4}
\int \frac{d^2b_t}{\pi} \,2\, [1 - e^{- \frac{1}{2} \sigma_N^{q \overline{q}} (x^{\prime}, \frac{4}{r_t^2}) S(b_t)}]\,\,.
\label{gm}
\end{eqnarray}
The lower limit in $r_t$ integration comes from the expression (\ref{ine}).

The space-time picture of the Glauber-Mueller approach is presented in 
Fig. \ref{fig1}. It takes into account only the interaction of the fastest 
partons with the target. As in QCD we expect that all partons from the 
cascade interact with the target, a generalized equation was proposed in 
\cite{ayala1}.

The AGL equation can be obtained directly from the Eq.(\ref{gm})  
differentiating this formula with respect to $ln \,1/x$ and 
$ln \, Q^2/\Lambda_{QCD}^2$. Therefore the AGL for the interaction of 
$q\overline{q}$ dipole is given by 
\begin{eqnarray}
\frac{\partial^2 xG_A(x,Q^2)}{\partial ln(1/x) \partial ln(Q^2/\Lambda_{QCD}^2)} = 
\frac{2\,Q^2}{\pi^2} \frac{C_A}{C_F}  
\int \frac{d^2b_t}{\pi}  \,[1 - e^{-\frac{1}{2}\sigma_N^{q\overline{q}}(x
,Q^2)S(b_t)}]\,\,, 
\label{agl}
\end{eqnarray}
where the dependence of  $\sigma_N^{q\overline{q}}$ in the  virtuality of the   
virtual probe results from the derivative. The nonperturbative effects 
coming from the large distances are absorbed in the boundary and initial 
conditions. This equation is valid in the double logarithmic approximation (DLA).

For a central collision ($b=0$) the AGL  equation (\ref{agl}) reduces to 
\begin{eqnarray}
\frac{\partial^2 xG_A(x,Q^2)}{\partial ln(1/x) \partial ln(Q^2/\Lambda_{QCD}^2)} = \frac{C_A}{C_F}
 \frac{2\,Q^2\,R^2}{\pi^2}  
  \,[1 - e^{-\frac{1}{2} \sigma_N^{q\overline{q}}(x
,Q^2)S(0)}]\,\,, 
\label{aglb0}
\end{eqnarray}
where we have set $db_t^2 = R^2$ for a $b=0$ collision. This limit is important since the cross section is strongly unitarized at small impact parameters \cite{ayala2, salam}. 

Considering that the  transverse cross-sectional area of the nucleus is $S_{\bot} = \pi  R^2$ and that $S(0) = A/(\pi R^2)$, the AGL equation for $b=0$ is obtained as
\begin{eqnarray}
\frac{\partial^2 xG_A(x,Q^2)}{\partial ln(1/x) \partial ln(Q^2/\Lambda_{QCD}^2)} = \frac{N_c \,C_F \, S_{\bot}}{\pi^3}
   Q^2[1 - e^{-\frac{2\alpha_s \pi^2 }{N_c S_{\bot}} \frac{1}{Q^2} xG_A(x,Q^2)}]\,\,, 
\label{aglkov}
\end{eqnarray}
where  $N_c = 3$  and  we have assumed that $C_F = N_c/2$ in the large $N_c$ limit. The Eq. (\ref{aglkov}) agrees with the expression (2) of \cite{kov}.

The AGL equation takes into account the interaction of all partons in a parton cascade with the target. In other words, the AGL equation takes into account that each parton in the parton cascade interacts with several nucleons within of the nucleus (Glauber multiple scattering).  In Fig. \ref{fig2} we present the result of the first iteration from the Glauber-Mueller formula, which is one of the terms summed by the AGL equation.

The GLR equation can be obtained directly from the Eq. (\ref{aglkov}). If we expand the right hand side of this equation to the second order in $xG_A$ we obtain
\begin{eqnarray}
\frac{\partial^2 xG_A(x,Q^2)}{\partial ln(1/x) \partial ln(Q^2/\Lambda_{QCD}^2)} = \frac{\alpha_s N_c}{\pi} \,xG_A(x,Q^2)
- \frac{\alpha_s^2 \pi}{S_{\bot}} \frac{1}{Q^2} [xG_A(x,Q^2)]^2 \,\,,
\label{glr}
\end{eqnarray}
which is the GLR equation for a cylindrical nucleus case (See Eq. (19) in \cite{kov}). Moreover, if the unitarity corrections are small, only the first order in $xG_A$ contributes. In this limit the equations (\ref{aglkov}) and (\ref{glr}) matches with the DGLAP evolution equation in the DLA limit.

Therefore the AGL equation 
(i)  matches   the DGLAP evolution equation 
in the DLA limit of PQCD; (ii) it 
has the GLR equation as a  limit, and (iii)  contains the 
Glauber-Mueller formula. We have that the AGL equation is valid in a 
large kinematic region.

Recently, a comprehensive phenomenological analysis of the behavior of distinct observables was made for the HERA kinematical region using the Glauber-Mueller approach \cite{ayala3,vic1,vic2}. In this kinematical region the solutions from the AGL equation and the Glauber-Mueller approximately coincide. 
The results from these analysis agree with the recent HERA data and allows to make some predictions which will be tested in a near future.
In \cite{vic1} we have analysed the behavior of the longitudinal structure function $F_L$ and the charm component of the proton structure function $F_2^c$ and have shown that our results agree with the H1 data. New data, with better statistics, will allow to  evidentiate the unitarity corrections. In \cite{vic2} we have  shown that the recent ZEUS data can be described if the unitarity corrections for the $F_2$ slope are considered. Our main conclusion is that the unitarity corrections cannot be disregarded in the HERA kinematical region.

\section{The Nuclear Structure Function}
\label{f2a}

The  unitarity corrections to  the nuclear structure function  can be   estimated in the rest frame of the target. This intuitive point of view was proposed by V. N. Gribov many years ago \cite{gribov}.  Gribov's assumption is that at small values of $x$  the virtual 
photon fluctuates into a $q\overline{q}$ pair well before the interaction 
with the target, and this system interacts with
the target. This formalism has been established as an useful tool for 
calculating deep inelastic and related diffractive cross section for 
$\gamma^*\,p$ scattering in the last years \cite{nik,buch}.
The Gribov factorization follows from the fact that the lifetime of the $q\overline{q}$
fluctuation is much larger than the time of the partonic interactions. According to 
the uncertainty principle, the fluctuation time is $\approx \frac{1}{m\,x}$, 
where $m$ denotes the target mass.

The space-time picture of the DIS in the target
rest frame can be viewed as the decay of the virtual photon at high energy
(small $x$) into a quark-antiquark pair long before the 
interaction with the target. The $q\overline{q}$ pair subsequently interacts 
with the target.  In the small $x$ region, where 
$x \ll \frac{1}{2mR}$, the $q\overline{q}$  pair 
crosses the target with fixed
transverse distance $r_t$ between the quarks. It allows to factorize the total 
cross section between the wave function of the photon and the interaction 
cross section of the quark-antiquark pair with the target. The photon wave function 
is calculable and the interaction cross section is modelled. Therefore, the 
nuclear structure function is given by 
\begin{eqnarray}
F_2^A(x,Q^2) = \frac{Q^2}{4 \pi \alpha_{em}} \int dz \int \frac{d^2r_t}{\pi} |\Psi(z,r_t)|^2 \, \sigma^{q\overline{q} + A}(z,r_t)\,\,,
\label{f2target}
\end{eqnarray}
where 
\begin{eqnarray}
|\Psi(z,r_t)|^2 = \frac{6 \alpha_{em}}{(2 \pi)^2} \sum^{n_f}_i e_f^2 \{[z^2 
+ (1-z)^2] \epsilon^2\, K_1(\epsilon r_t)^2 + m_f^2\, K_0^2(\epsilon r_t)^2\}\,\,,
\label{wave}
\end{eqnarray}
$\alpha_{em}$ is the electromagnetic coupling constant,
$\epsilon = z(1-z)Q^2 + m_f^2$, $m_f$ is the quark mass, $n_f$ is the number 
of active flavors, $e_f^2$ is the square of the  parton charge (in units of $e$), $K_{0,1}$ 
are the modified Bessel functions and $z$ is the fraction of the photon's light-cone 
momentum carried by one of the quarks of the pair.  In the 
leading log$(1/x)$ approximation we can neglect the change of $z$ during the 
interaction and describe the cross section $\sigma^{q\overline{q}+A}(z,r_t^2)$ as 
a function of the variable $x$.

We estimated the unitarity corrections considering the  Glauber multiple scattering theory \cite{chou}, 
which was probed in QCD \cite{muegla}. The nuclear
 collision is analysed as a
sucession of collisions of the probe with individual nucleons within the nucleus, which implies that   
the $F_2$ structure function can be written  as \cite{ayala1}
\begin{eqnarray}
F_2^A(x,Q^2) =  \frac{Q^2}{4 \pi \alpha_{em}} \int dz \int \frac{d^2r_t}{\pi} |\Psi(z,r_t)|^2 \,  \int \frac{d^2b_t}{\pi}  
\,2\,\{1 - e^{-\frac{1}{2}\Omega_{q\overline{q}}(x,r_t,b_t)}\}\
\label{f2eik}
\end{eqnarray}
where the opacity $\Omega_{q\overline{q}}(x,r_t,b_t)$ describes the interaction 
of the $q\overline{q}$ pair with the target.

In the region where  $\Omega_{q\overline{q}}$ is small $(\Omega_{q\overline{q}} \ll 1)$ the  
$b_t$ dependence can be factorized as $\Omega_{q\overline{q}} = \overline{\Omega_{q\overline{q}}} S(b_t)$ \cite{100}, 
with the normalization $\int d^2b_t\, S(b_t) = 1$.  The eikonal approach assumes that 
the factorization of the $b_t$ dependence 
$\Omega_{q\overline{q}} = \overline{\Omega_{q\overline{q}}} S(b_t)$, which is 
valid in  the region where  $\Omega_{q\overline{q}}$ is small, occurs in the whole kinematical region \cite{ayala2}.
The main assumption of the eikonal approach in pQCD is the identification of opacity
$\overline{\Omega_{q\overline{q}}}$ with the gluon distribution.
In \cite{plb} the opacity  is given by $\overline{\Omega_{q\overline{q}}} = \sigma_N^{q\overline{q}}$.

For a central collision ($b=0$) the Eq. (\ref{f2eik}) reduces to 
\begin{eqnarray}
F_2^A(x,Q^2) =  \frac{Q^2}{4 \pi \alpha_{em}} R^2 \int dz \int \frac{d^2r_t}{\pi} |\Psi(z,r_t)|^2 \,   
\,2\,\{1 - e^{-\frac{1}{2}\sigma_N^{q\overline{q}}S(0)}\}\,\,.
\label{f2eik2}
\end{eqnarray}
Substituting $\sigma_N^{q\overline{q}}$ and $S(0)$ we obtain that the nuclear structure function for $b=0$ is given by
\begin{eqnarray}
F_2^A(x,Q^2) =  \frac{Q^2}{4 \pi \alpha_{em}} R^2 \int dz \int \frac{d^2r_t}{\pi} |\Psi(z,r_t)|^2 \,   
\,2\,\{1 - e^{-\frac{ \alpha_s C_F \pi^2}{N_c^2 S_{\bot}}r_t^2 AxG(x,1/r_t^2)}\}\,\,.
\label{f2eik3}
\end{eqnarray}

This expression estimates the unitarity corrections to the nuclear structure function for central collisions ($b=0$) in the DLA limit using the Glauber-Mueller approach.

Comparing the Eq. (\ref{f2eik3}) with the expression to the nuclear structure function in the dipole picture  proposed  in \cite{kov}, we obtain the  representation for the total cross section of the $q \overline{q}$ pair interacting with the nucleus, $N(\vec{x_{01}}, \vec{b_0} = 0, Y)$, used in that reference,
\begin{eqnarray}
N(\vec{x_{01}}, \vec{b_0} = 0, Y) = \,2\,\{1 - e^{-\frac{ \alpha_s C_F \pi^2}{N_c^2 S_{\bot}} x_{01}^2 AxG(x,1/x_{01}^2)}\}\,\,,
\label{ene}
\end{eqnarray}
where $x_{01} = x_0 - x_1 \equiv r_t$ [$x_0$ ($x_1$) is the transverse coordinate of the quark (antiquark)] and $Y=ln\,(s/Q^2) = ln\,(1/x)$. This relation between $N(\vec{x_{01}}, \vec{b_0} = 0, Y)$ and the gluon distribution is valid in the DLA limit and  is two times the propagator of the $q\overline{q}$ pair through the nucleus obtained in \cite{kovmue} at large $Q^2$ (See Eq. (6b) of \cite{kov}). 

We will assume that the correct expression for $N$ in the DLA limit is given by (\ref{ene}). In the next section we will use this expression as an input in the evolution equation for $N$  obtained in the dipole picture.

\section{The AGL Equation from the Dipole Picture}
\label{agldip}

Considering the multiple pomeron exchange, Kovchegov \cite{kov} has obtained an evolution equation for  $N(\vec{x_{01}}, \vec{b_0}, Y)$ in the leading logarithmic approximation
\begin{eqnarray}
N(\vec{x_{01}}, \vec{b_0}, Y) = - \gamma (\vec{x_{01}}, \vec{b_0}) exp\,\left[-\frac{4 \alpha_s C_F}{\pi} ln \left(\frac{x_{01}}{\rho}\right)\,Y\right] \nonumber \\
 + \frac{\alpha_s C_F}{\pi^2} \int_0^Y dy \, exp \, \left[-\frac{4 \alpha_s C_F}{\pi} ln \left(\frac{x_{01}}{\rho}\right)\,(Y-y)\right]\,  \nonumber \\
\times \int_{\rho} d^2x_2 \frac{x_{01}^2}{x_{02}^2 x_{12}^2} \,  [2N(\vec{x_{02}}, \vec{b_0}, y)  
 -  N(\vec{x_{02}}, \vec{b_0}, y)N(\vec{x_{12}}, \vec{b_0}, y)]\,\,,
\label{kovllx}
\end{eqnarray}
where $x_{ij} = x_i - x_j$ is the size of the dipole with a quark in the transverse coordinate $x_i$ and the antiquark in $x_j$, $\gamma (\vec{x_{01}}, \vec{b_0})$ is the propagator of the $q\overline{q}$ pair through the nucleus and  $\rho$ is an ultraviolet cutoff in the equation which disappears in the physical quantities. We have reobtained  this equation and denote it  as the K equation.

The equation (\ref{kovllx}) was obtained considering the scattering of a virtual photon with a nucleus. The physical picture for this interaction is the same as the Glauber-Mueller approach. The incoming virtual photon generates a  $q\overline{q}$ pair which develops a cascade of gluons, which then scatters on the nucleus. In the large $N_c$ limit the gluon can be represented as a $q\overline{q}$ pair. Therefore, in this limit  and in the leading logarithmic approximation, the cascade of gluons can be interpreted as a dipole cascade, where each dipole in the cascade interacts with several nucleons within  the nucleus. Therefore, as the  K equation  and the AGL equation, although with distinct basic objects, resums the multiple scatterings of its respective degree of freedom, we expect that both coincide in a common limit.

In the double logarithmic limit, where the momentum scale of the photon $Q^2$ is larger than the momentum scale of the nucleus $\Lambda_{QCD}$,  we take the large $Q^2$ limit from  (\ref{kovllx}), which  reduces to
\begin{eqnarray}
\frac{\partial N(\vec{x_{01}}, \vec{b_0}, Y)}{\partial Y} & = & \frac{\alpha_s C_F}{\pi} x_{01}^2 \int_{x_{01}^2}^{1/\Lambda_{QCD}^2} \frac{d x_{02}^2}{(x_{02}^2)^2} [2N(\vec{x_{02}}, \vec{b_0}, Y)  \nonumber \\ 
& - & N(\vec{x_{02}}, \vec{b_0}, Y)N(\vec{x_{02}}, \vec{b_0}, Y)]\,\,.
\label{kovdla}
\end{eqnarray}
This equation considers the evolution of the dipole  transverse sizes from a small scale $x_{01}$ to a large scale $1/\Lambda_{QCD}$.
Differentiating the above expression with respect to $ln \,(1/x_{01}^2 \Lambda_{QCD}^2)$  a double differential  equation for $N(\vec{x_{01}}, \vec{b_0}, Y)$ follows
\begin{eqnarray}
\frac{\partial^2 N(\vec{x_{01}}, \vec{b_0}, Y)}{\partial Y \partial ln \,(1/x_{01}^2 \Lambda_{QCD}^2)} = 
\frac{\alpha_s C_F}{\pi}\,[2  -  N(\vec{x_{01}}, \vec{b_0}, Y)] N(\vec{x_{01}}, \vec{b_0}, Y)\,\,.
\label{kovdif}
\end{eqnarray}
The physical picture for the dipole evolution in the DLA limit is that the produced dipoles at each step of the evolution have a much greater transverse dimensions that the parent dipoles .

In \cite{kov} the connection between the quantity  $N(\vec{x_{01}}, \vec{b_0}, Y)$ and the nuclear gluon distribution was discussed.  As there is some freedom in the definition of the gluon distribution, the choice for the connection between the two functions was arbitrary. Here we use the result obtained 
 in the previous section [Eq. (\ref{ene})] to make more explicit this connection. If  the Eq. (\ref{ene}) is expanded to the first order in $xG_A$, our result differs from the choice used in \cite{kov}  by a factor of 2.

Substituting Eq. (\ref{ene}) into Eq. (\ref{kovdif}) and using $x_{01} \approx 2/Q$ \cite{kov}, which is valid in the double logarithmic limit, we end up with 
 \begin{eqnarray}
\frac{\partial^2 xG_A(x,Q^2)}{\partial ln(1/x) \partial ln(Q^2/\Lambda_{QCD}^2)} = \frac{N_c \,C_F \, S_{\bot}}{\pi^3}
   Q^2[1 - e^{-\frac{2\alpha_s \pi^2 }{N_c S_{\bot}} \frac{1}{Q^2} xG_A(x,Q^2)}]\,\,, 
\label{aglkov2}
\end{eqnarray}
which exactly corresponds to the AGL equation [Eq. (\ref{aglkov})]. As demonstrated in the Section \ref{agls} the GLR equation is a straithforward consequence from the AGL equation.

A last comment is important. Using the relation between $N$ and $xG$ from \cite{kov}  a factor $e^{ -\frac{\alpha_s \pi^2 x_{01}^2 xG_A}{N_c S_{\bot}}}$  remains in the right-hand side from the Eq. (\ref{aglkov2}). Therefore, the AGL equation only could be justified for a configuration of very small dipoles, as was discussed in \cite{kov}. 
Consequently, the connection between the dipole and AGL approaches deals on the correct physical relation between the function $N$ and the nuclear gluon distribution, given here by Eq. (\ref{ene}).

\section{Conclusions}

The AGL equation, obtained using the Glauber-Mueller approach, resums all multiple pomeron exchanges in the double logarithmic limit. Its solution allows to make predictions which agree with the recent HERA data.
In this paper we have analysed another equation which includes all multiple pomeron exchanges proposed by Kovchegov recently. The K equation  was obtained in the leading logarithmic approximation using the dipole picture. We agree with this result. However, as the relation between    the function $N$, which  evolution is described by the K equation,  and the nuclear gluon distribution has some freedom, we obtained this connection considering the nuclear structure function in the Glauber-Mueller approach. Substituting this relation in the evolution equation for $N$ we have shown that it reduces directly to the AGL equation and, as a limit, to  the GLR equation in the double logarithmic limit.  
This result shows that the AGL equation is a good candidate for the unitarized evolution equation at small $x$ in the DLA limit, supported by two different frameworks describing small $x$ phenomena.

Another candidate for the unitarized evolution equation was proposed by Jalilian-Marian {\it et al.} \cite{jamal}. These authors have derived a general evolution equation for the gluon distribution  in the limit of large parton densities and leading logarithmic approximation, considering  a very large nucleus. This work is based on  effective Lagrangian formalism for the low $x$ DIS \cite{mcl} and the Wilson renormalization group.
 In the general case the evolution equation  is a very complicated equation, which does not allow to obtain analytical solutions. Recently, these authors have considered the DLA limit on their result \cite{jamal1} and have shown 
that the evolution equation reduces to an equation with a functional form similar, but no identical, to the AGL equation. We believe that a more detailed analysis of the approximations used in both equations will allow to  demonstrate the equivalence of both equations in a common limit  \cite{lev}.

\section*{Acknowledgments}

This work was partially financed by CNPq and by Programa de Apoio a N\'ucleos de Excel\^encia (PRONEX), BRAZIL.

\newpage

\begin{figure}
\caption{Space-time  picture of the Glauber-Mueller approach. The  
virtual colorless hard probe  $G^*$ decays into a pair with transverse 
distance $r_t$  which interacts with the $n$ nucleons within  the nucleus.  }
\label{fig1}
\end{figure}

\begin{figure}
\caption{First iteration of the Glauber-Mueller approach. The AGL equation 
takes into account the interation of all partons  of the parton cascade  
with several nucleons within of the nucleus.}
\label{fig2}
\end{figure}

\end{document}